# Cross-correlated TIRF/AFM shows Self-assembled Synthetic Myosin Filaments are Asymmetric – Implications for Motile Filaments

André E.X. Brown*,[#], Alina Hategan[§], Daniel Safer[§], Yale E. Goldman[#,§,¶], Dennis E. Discher[#,¶,‖]

*Department of Physics and Astronomy, [#]Nano/Bio Interface Center, [§]Pennsylvania Muscle Institute, and Graduate Groups in [¶]Cell and Molecular Biology and [‖]Physics,
University of Pennsylvania, Philadelphia, PA, 19104
André E.X. Brown and Alina Hategan contributed equally to this work.

ABSTRACT  Myosin-II's rod-like tail drives filament assembly with a head arrangement that should generate equal and opposite contractile forces on actin – if one assumes that the filament is a *symmetric* bipole.  Self-assembled myosin filaments are shown here to be asymmetric in physiological buffer based on cross-correlated images from both atomic force microscopy (AFM) and total internal reflection fluorescence (TIRF).  Quantitative cross-correlation of these orthogonal methods produces structural information unavailable to either method alone in showing that fluorescence intensity along the filament length is proportional to height.  This implies that myosin heads form a shell around the filament axis, consistent with F-actin binding.  A motor density of ~50-100 heads/micron is further estimated but with an average of 32% more motors on one half of any given filament compared to the other, regardless of length. A purely entropic pyramidal lattice model is developed that qualitatively captures this lack of length dependence and the distribution of filament asymmetries.  Such strongly asymmetric bipoles are likely to produce an imbalanced contractile force in cells and in actin-myosin gels, and thereby contribute to motility as well as cytoskeletal tension.
(186 words)

**Introduction**

Myosin-II molecules are found in contractile 'bipolar' filaments in cell types that range from striated muscle to stem cells (1). A number of these myosin-IIs are also known to self-assemble *in vitro* into active bipolar filaments, and both the structure and assembly characteristics of such 'synthetic' filaments have been intensively studied for several decades—including recent work with AFM (2,3)—with most studies focusing on skeletal muscle myosin (4). Synthetic filaments have also found use in *in vitro* motility assays where it has been shown that myosin arranged into ordered filaments moves actin filaments differently from single myosins randomly oriented on a surface (5-7). More recently, synthetic filaments have also been incorporated into cross-linked actin networks to make "active gels" that might be considered *in vitro* mimics of cellular cytoskeletal systems (8). Because of the long interest in using synthetic filaments as model systems to study the motility and contractility of myosin, further insight into filament properties could be important. Myosin filaments already have many well-characterized structural features and are therefore an excellent sample to develop 'Cross-correlated TIRF/AFM' in which two orthogonal imaging methods – based respectively on optics and topography – are mathematically compared and fit to a model. This provides the dual advantage of testing the reliability of the coupled method in a well-defined system while also opening up the possibility of new insights into these important motor filaments.

The model of thick filament structure first suggested by Huxley in 1963 (9) proposed the 150 nm long rod domain of myosin assembles in the core of the filament with the heads decorating the exterior of the filament – this model has stood the test of time. Many subsequent studies of myosin filament structure have focused on the mechanisms governing the filament length distribution, which under certain conditions can be quite close to the physiological length of 1.5 µm (10). The narrow length distribution is not the only feature that is important for muscle myosin filaments *in vivo*: they are found to be highly symmetric, which seems appropriate for muscle contraction (11). However, while synthetic filaments have been shown to be bipolar in motility assays (6,7), their symmetry has not been sufficiently addressed before. In this paper, we combine single molecule fluorescence in TIRF microscopy with AFM tapping mode imaging to study the arrangement of myosin heads in synthetic filaments, showing that their symmetry is in fact not strictly controlled.

**Materials and Methods**

*Myosin purification and fluorescence labeling*    Myosin was prepared from rabbit skeletal muscle as described by Margossian & Lowey (12). Myosin molecules were covalently labeled with tetramethylrhodamine-5-maleimide (Molecular Probes, Eugene, OR) or tetramethylrhodamine-5-iodoacetamide (AnaSpec, San Jose, CA), under conditions that favor labeling at Cys-707 in the myosin head (13) at ~1:1 stoichiometry. Myosin in 40 mM KCl, 5 mM sodium phosphate, 0.1 mM tris(2-carboxyethyl)phosphine, pH 7.0, was incubated at 4$^{o}$C for 16 h with either tetramethylrhodamine-5-maleimide or tetramethylrhodamine-5-iodoacetamide, at a ratio of 1:1. The labeled myosin was recovered by centrifugation at 12,000g for 10 min and was then redissolved in 0.6 M KCl, 10 mM imidazole-HCl, 2 mM MgCl$_2$, 1 mM dithiothreitol, pH 7.0, and dialyzed against the same buffer. The myosin stock had a concentration of 3.13 mg/ml and was stored in 50 % glycerol solution at –20$^{o}$C.

*Myosin filament preparation*     Myosin from the stock was diluted 10 times in 300 mM KCl, 5 mM MgCl$_2$, 10 mM Hepes buffer at pH 7, and centrifuged at 100 000 rpm for 30 min at 4$^0$C to sediment any aggregates leaving single myosin molecules in the supernatant. Filament preparation was done by rapid mixing of myosin monomers in the supernatant solution with an equal volume of polymerization buffer (5 mM MgCl$_2$, 10 mM Hepes buffer at pH 7) or by dialysis in which the polymerization buffer was added in 15 steps at 1 min intervals, during continuous mixing of the sample. Samples were incubated for 30 min at room temperature before storing on ice. Experiments were performed on the same day the filaments were formed.

*Sample preparation for TIRF/AFM*     For AFM imaging, filaments from the polymerization vial were diluted 10 times in 150 mM KCl, 5 mM MgCl$_2$, 10 mM Hepes buffer at pH 7. 1 μl of 1 mg/ml BSA was added to 100 μl of this solution. 50 μl of this solution was then allowed to adhere to clean glass slides that were spin-coated with poly(methyl methacrylate) (PMMA) (Sigma-Aldrich, Saint Louis, MI, catalogue number: 370037) for 1 min. The sample was then gently washed with the same buffer used for dilution.

*Hybrid TIRF/AFM*     A Veeco Bioscope II AFM mounted on a Nikon TE 200 inverted optical microscope was equipped with an Olympus oil immersion TIRF objective (60x magnification, 1.45 numerical aperture). Fluorescence was excited using a 300 mW 532 nm laser (B&W TEK, Newark, DE) whose intensity was adjusted by rotating one of two crossed polarizers to minimize photobleaching while still allowing for single molecule imaging with a reasonable frame rate. The beam was then directed into the back of the microscope from a mirror with an adjustable angle and converging lens used to control the position of the beam within the objective and therefore the angle of incidence at the sample coverslip. The incident angle was set to obtain strong reflection from the PMMA water interface without cutting off the intensity at the objective back focal plane aperture. Using an AFM tip coated with fluorescent protein, we measured the spatial decay length of the evanescent field to be 140 nm (Appendix A), corresponding to an incident angle in the glass of 63 degrees.

Fluorescence images were collected on a 16 bit Cascade 512B EMCCD camera (Photometrics, Tucson, AZ) with a frame rate of 50 ms. Each image was 512 by 512 pixels with no additional binning.

*AFM Imaging*     All AFM imaging was done in buffer in tapping mode using either DNP-S or MSCT silicon nitride cantilevers (Veeco, Santa Barbara, CA). Images were 512 by 512 pixels large and were taken with a variety of tip velocities, usually around 10 μm/s. AFM images were taken at regions of interest determined from previously captured TIRF images based on the known tip location as determined from brightfield imaging of the cantilever.

*Image Analysis*     Registration was done manually based on the known region of interest determined from the TIRF image before AFM imaging. Height and intensity profiles were measured along the center line of the filaments using NIH ImageJ after the filaments had been straightened using the Straighten plugin by Kocsis et al. (*14*). The profiles show the intensity at each point along the length without any averaging or interpolation between pixels. All other analysis was performed using Mathematica (Wolfram Research, Champagne, IL). The data for the height-intensity scaling plot (Fig. 3a) were taken from the filaments shown in Fig. 2 and two others from a different region that were long enough for profiles to be determined from the TIRF image and that were also well separated and in regions with low background fluorescence.

**Results and Discussion**
*AFM vs. TIRF Resolution*

Single molecule imaging by fluorescence has been achieved with a large number of systems, but always when individual fluorophores were separated by a distance greater than the diffraction limit (*15*) or were activated or bleached sequentially and thus effectively separated in time (*16*). In filaments, myosin molecules and therefore fluorophores are crowded into a tight space, well beyond the diffraction limit of a light microscope. Indeed, when formed by rapid mixing, filaments proved too short for any structural information to be determined from the TIRF images alone (**Fig. 1a**), but the bright spots visible in TIRF clearly show the presence of spots that fluoresce too intensely to be single molecules (per intensity analyses below). However, when the same region is imaged with AFM, the pixelated spots are resolved into filaments with the clearly tapered ends expected for myosin filaments prepared under these conditions (**Fig 1b**) (*17*). Lengths can also be accurately determined as $0.9 \pm 0.2$ μm (mean ± S.D.), which is somewhat shorter than the physiological length of 1.5 μm (*9*).

Given the significant resolution advantage offered by AFM, it might not be immediately clear that the two images do in fact show the same region of the sample. To aid in comparison, the TIRF image was thresholded and the outline of this thresholded image is shown as the white overlay in Fig. 1b. It should now be clear that both images show the same sample region and that the brightest regions in TIRF are in fact due to several filaments that cannot be resolved optically. The AFM used has a closed-loop position sensor that corrects possible errors from piezo non-linearity, hysteresis, and drift and allows specific regions to be imaged at still higher resolution in AFM in order to look for more detailed structural features of the samples than is possible in fluorescence (compare **Fig. 1c, d**). At this resolution, the tapered structure becomes more apparent and it is clear that there is no central bare zone—a region without myosin heads—as expected for pH 7 filaments (*17*).

*Quantitative Cross-correlation of TIRF and AFM Data*

Longer myosin filaments with a broader length distribution $3 \pm 2$ μm (mean ± S.D.) are formed when dialysis is used to slowly reduce the ionic strength of a solution of myosin monomers (Materials and Methods). Filaments formed using this method have proven useful for *in vitro* motility assays because actin movement can be tracked over longer distances that can be clearly observed in optical microscopy (*6*). In the case of combined TIRF/AFM, the fact that longer filaments can be clearly resolved in TIRF makes a quantitative comparison of fluorescence and height data possible. **Fig. 2a** shows a TIRF image of three myosin filaments, and **Fig. 2b** shows the AFM height image of the same region. The images are clearly similar, and quantitative analyses add further insight.

Intensity and height profiles taken along the filament centerlines are plotted in **Fig. 2c** (fluorescence intensity in red and height in black). To facilitate comparisons by taking the broadening of the optical microscope into account, we smoothed the AFM height profiles by convolution with a Gaussian kernel with the same standard deviation as that determined from fits to the intensity profiles of single fluorophores (Fig. 2a, inset). The smoothed height profile is shown in blue.

The agreement between the two methods is clear upon inspection, but to make these observations more precise, we made a height-intensity scatter plot using the profiles in Fig. 2c as well as two from another sample region. The intensity is fit with a function of the form $I = H^a + c$ with $a = 1.0 \pm 0.1$ suggesting that the scaling is linear. To confirm that the data are in fact represented better by a linear model, we used the Akaike information criterion (18,19). This is useful because simply considering the goodness of fit will favor models with more parameters. In the extreme case of a polynomial with the same degree as the number of data points, the fit will always be perfect. In contrast, the Akaike information criterion balances this increase in goodness of fit with a term that depends on the number of parameters. The results of this analysis can be converted to evidence ratios (*19*) that revealed a linear fit is 3-fold more likely to account for the data than a power law fit and is $3 \times 10^6$ times more likely than a quadratic fit. This linear scaling is not obvious *a priori* and in fact reveals information about filament structure that is not available from either TIRF or AFM alone. In particular, because height simply reflects the filament diameter whereas intensity reflects the total number of fluorophores in a filament cross-section, linear scaling implies that the fluorophores, which reside in the myosin head, are arranged in a shell around the filament. In other words, if the heads are arranged in a shell around the filament they will form a circle in cross-section. If their diameter is doubled, the circumference of this circle and therefore the number of myosin heads will also double. In contrast, had the labeling been uniform throughout the filament, a quadratic scaling would be expected since in this case the intensity would scale as the filament cross sectional area while the height would still reflect the filament diameter. An intermediate scenario with some labeled heads in the filament interior might also have been possible, but this is 3-fold less likely to account for the observed data as determined by the Akaike information criterion. These possibilities are displayed schematically in **Fig. 3b**.

The analyses above could be complicated by several factors. The first is that the evanescent field used to excite fluorescence in TIRF decays exponentially from the sample surface so that heads on the top of a filament will experience a smaller excitation intensity than those on the surface. However, this correction can be shown to be negligible based on reasonable assumptions about the parameters of this experiment and a direct measurement of the fluorescence intensity as a function of height (see Appendix A). Another potential source of error comes from inhomogeneities in the excitation across the field of view that distort the intensity along the filament length. Based on images from surfaces densely covered with fluorophores, these variations are not significant over the length scale of the filaments. Finally, it is also possible that the filament is deformed either by adsorption to the surface or from compression by the AFM tip. To minimize this effect, care was taken to keep tip-sample forces as small as possible by adjusting the amplitude set point as close to the free amplitude as possible while still maintaining surface contact and image quality.

Consistent with past work on filaments formed under the pH 7 conditions used here, there appears to be an 'adventitious' surface layer of myosin molecules that would explain the lack of a visible central bare zone (20,21). Non-physiological conditions can sometimes enhance the bare zone (*17*), and for these filaments molecular exchange of the inner-most molecules with the surrounding medium is estimated to be negligible (*22*). The shell-like arrangement of myosin heads in 'pH 7 filaments', sets some limits on the adventitious layer: the density of the total shell of heads – including both the adventitious layer and the core – is roughly constant for all filament heights, and therefore along the filament length.

*Myosin Numbers from AFM and Single Molecule TIRF Imaging*

Another test of the agreement of the two imaging methods is an estimate of the number of molecules in the filament based on each method. Since the structure of myosin is known, we can estimate the volume per molecule and compare it to the total filament volume. Each myosin molecule is modeled as a 150 nm long rod with a 2 nm diameter and two heads that are each 10x5x3 nm for a total volume per myosin of ~770 nm$^3$. Measuring volume from AFM images requires some knowledge of either the sample or tip shape. The tips we used in this case have a nominal initial radius of curvature at the tip between 10 and 40 nm, however this can increase during scanning due to blunting against the surface or adsorption of material to the tip. The filament cross-sectional height profile shown in the inset in Fig. 2B has a width at its base more than ten-fold greater than the height of 5 nm, implying that some broadening due to the tip/sample convolution has indeed occurred or that the filament flattened somewhat during adsorption to the surface. In principle the contributions of broadening and flattening could be assessed by performing a deconvolution using some assumption for the tip geometry, but because the filaments are only a few pixels across, this does not seem justified. Instead, we have chosen to assume the filament has a circular cross-section with a diameter equal to the filament height at that point, and to calculate the volume based on this assumption. For this reason, our volume estimates from AFM should be taken as lower limits. **Table 1** lists the results for five filaments that range from about 50 to 450 myosin motors per filament.

To estimate the number of molecules based on the TIRF data we followed an analogous procedure and first determined the average intensity of single, well-separated myosin molecules. We imaged at high ionic strength (0.6 M KCl, 10 mM imidazole-HCl, 2 mM MgCl$_2$) ensuring that no filaments were formed, and then sought spots that bleached in one step, fitting them with a 2-dimensional Gaussian (Appendix B). This procedure yields an average single molecule intensity and was repeated each day immediately before filament experiments. The number of myosin motors per filament obtained from TIRF (Table 1) was thus obtained by comparing the average single molecule intensity to the total intensity of a filament, and the estimates appear similar in magnitude to those obtained with AFM. Indeed, a plot of $N_{TIRF}$ = (Myosins/Filament)$_{TIRF}$ versus $N_{AFM}$ = (Myosins/Filament)$_{AFM}$ can be fit to a line: $N_{TIRF} = 11 + 0.67 N_{AFM}$ ($R^2$ = 0.98). With TIRF, photobleaching of the fluorophores in the filament is unavoidable and could explain the persistent underestimate by ~33%. Depending on the arrangement of heads in the filaments, it is possible that fluorophore self-quenching could also contribute to this underestimate. The Förster radius for rhodamine self-quenching by resonant energy transfer is likely between 4-5 nm (the Förster radius for homo-FRET of fluorescein, which has a similar overlap between its excitation and emission spectra, is 4.0 nm (*23*)). Based on the number of myosins per µm (see below) the average inter-head distance is between 10 and 20 nm. Because of the rapid decay of the transfer efficiency with distance, the decrease in intensity due to self-quenching should not be significant. In either case, a linear model provides an excellent fit in comparing AFM and TIRF.

Our determinations of motor numbers $N_{TIRF}$ and $N_{AFM}$ are in sufficiently good agreement with each other that we can estimate a mean motor density of $\rho_{myo} \sim 50\text{-}100$ myosins per μm. This estimate for $\rho_{myo}$ is about 25-50% of the physiological value in muscle of 210 molecules per μm (each crown of three molecules is separated by 14.3 nm (*24*)), and it is smaller still than an estimate based on the spacing of single fluorophores in sparsely labeled, but otherwise similarly prepared filaments (*25*). In our experiments, a 30% decrease in filament heights due to adsorption and/or compression under the tip would give a 50% underestimate for $\rho_{myo}$, and such dimensions (≤2.5 nm decrease) are small and very possible when compared to myosin head and tail dimensions.

*Filament Asymmetry $\rho_{myo}(s)$*

Since the fluorescent labels reside primarily in the myosin head and since the correlation of the data from our two modalities leads to a reasonable model of head arrangement as well as mutual agreement about $\rho_{myo}$, we can reasonably assume that the height variations seen along the filament contour *s* in AFM traces reflect the local density of myosin heads. The profiles for $\rho_{myo}(s)$ plotted in Fig. 2c show the filaments are generally not symmetric about their midpoints. The fractional asymmetry δ is defined simply as the volume of one half of the filament minus the volume of the other half divided by the total volume. The volumes are determined from the height profile as described above with the midpoint determined from the total filament length.

The results of this analysis are shown as inset numbers in Fig. 2c and the asymmetries measured from a larger set of filaments are summarized in **Fig. 4**, which includes data from filaments formed by both rapid mixing and by dialysis with both showing similar values for the asymmetry. There is also no clear trend in the fractional asymmetry with length, highlighting the fact that even the shorter filaments formed by rapid mixing do not have physiological symmetry. The probability distribution $P(\delta)$ decreases almost linearly with increasing δ so that symmetric filaments are the most likely, but since the decay is not very rapid the mean $<P(\delta)>$ is still 0.32.

*A Purely Entropic Lattice Model of Asymmetric Filaments*

The lack of length dependence on the asymmetry as well as the decrease of $P(\delta)$ with δ can be captured using a purely entropic lattice model for the filaments. The brick-wall lattice with no overhangs shown in **Fig. 5a** is motivated by the staggered packing of myosin molecules in filaments and has the desirable feature that filaments taper at their ends, as observed in experiments. By taking the binding energy to be constant for all sites, the statistical mechanical problem is reduced to a combinatorial problem: a purely entropic description of the equilibrium in this model thus depends only on the number of possible arrangements of molecules on the pyramidal lattice. Counting the number of ways of arranging bricks on this lattice is equivalent to counting the number of Dyck paths of the same size, i.e. the number of ways of connecting the two corners of a square region on a square lattice without crossing the line $y = x$. This is a useful equivalence to note (shown schematically in Fig. 5a), because the solution of this second problem is known. The number of possible Dyck paths with base length $L$ is given by the Catalan numbers (*26*)

$$C_L = \frac{(2L)!}{(L+1)!L!} \quad \text{for } L \geq 0.$$

To our knowledge, no analytical expression for the fractional asymmetry of these states is known but it is possible to generate all possible Dyck paths for a given base length and to then calculate the fractional asymmetry $\delta$ for each of these model filaments.

Although $C_L$ grows very rapidly with $L$, the results prove sufficiently informative for $L$ up to 11 as shown in **Fig. 5b**. $\delta$ is zero for filaments of length $L = 1$ and 2, but this rises and rapidly plateaus reaching a nearly constant value. Thus the pyramidal lattice model qualitatively captures the length dependence of $\delta$ shown in Fig. 4. The asymptotic value of 0.14 is about half the experimental value, but it is important to note that these planar model filaments with a base length of 11 can have up to 66 molecules in them so that the size ranges explored in the model are comparable to the shorter filaments observed in AFM.

The form of $P(\delta)$ predicted by the model (**Fig. 5c**) is also qualitatively similar to the experimental distribution: small $\delta$ is the most likely but most of the filaments are asymmetric. One difference between the distribution predicted by the model and that observed in AFM is the range of the data. The maximum asymmetry predicted for filaments of length $L = 11$ is 0.5, whereas experimental values occasionally reach above 0.7. This is partly because of the limited length $L$ that is accessible, since as is clear from the plot in Fig. 5b the upper range of asymmetries is still increasing even if the average is not. Nonetheless, given the simplicity of the pyramidal lattice model considered here, the qualitative agreement suggests that even a slightly more sophisticated model of synthetic filaments might capture more details of the filament structure and perhaps also give insight into the kinds of regulatory mechanisms that control filament symmetry *in vivo*.

*Asymmetry Implies a Force Imbalance*

Since these asymmetries are correlated with the number of heads on either half of the filament, asymmetric filaments imply a force imbalance even if the filaments are bipolar and are attached to oppositely oriented actin filaments, as shown schematically in **Fig. 6**. In other words, in a cross-linked network of actin filaments, as used in active gel experiments (*8*), where the actin network is relatively immobile, myosin filaments might not be simply contractile, but in fact motile, because even at a junction, asymmetric filaments should move towards the end with more heads. Furthermore, on a single filament or bundle of parallel actin filaments with the same polarity, even symmetric filaments are expected to be motile given the results of actin gliding assays that show that actin can travel in both directions along both parts of myosin filaments, albeit more slowly in the non-physiological direction (*6,7*). This kind of motility could be important for the comparison between models of active gels and quantitative data on contractility (*27*). In future experiments, the potential motility of myosin in active gels could be determined in a straightforward way by using different color labels for myosin and actin.

In muscle, filament length and symmetry are tightly controlled but less is known about the structure of stress fibers in non-muscle cells. While there are several proteins that are involved, their role in controlling myosin filament symmetry is unknown, and slight asymmetries in myosin mini-filaments could be particularly important in non-muscle cells. It has already been proposed that myosin filaments could be motile—rather than purely contractile—in stress fibers with uniform polarity (*28*) but this may even be the case in sarcomere-like stress fibers if there is a force imbalance due to asymmetric myosin filaments. This kind of motility could serve simply to transport myosin or could perhaps be involved in cargo transport as other myosins are known to be (*29*). In any case, if filament motility is absent in non-muscle cells this may imply that there is a mechanism regulating non-muscle myosin filament symmetry or some other mechanism to stabilize asymmetric filaments in actin bundles or junctions.

**Conclusions**

Cross-correlated-TIRF/AFM provides complementary advantages for sample characterization: TIRF can leverage the extensive chromophore and fluorescent protein technology that has been developed to specifically identify molecular species of interest, while AFM can be used to collect high resolution structural information from the same sample region. Furthermore, a quantitative correlation of the data from the two modalities provides structural information that would have been unavailable from either one alone. In particular, we were able to confirm that the heads of myosin filaments are arranged in a shell of roughly constant thickness around the filament, consistent with having the heads exposed to the outside as required for their actin binding function. It also shows the power of correlative TIRF/AFM to determine the arrangement of fluorophores *within* a structure that is only nanometers thick in fluid at room temperature. The same method can be extended to other macromolecular structures that can be fluorescently labeled and deposited on a surface.

Using the structural information revealed by AFM images of myosin filaments, we quantified the fractional asymmetry of synthetic filaments and found that it averaged 0.32 for filaments prepared by rapid mixing and by dialysis and over the length range we observed. This asymmetry could have implications for intracellular trafficking, (*28*) *in vitro* motility measurements (7) and for experiments using myosin to activate actin gel systems (*27*).

**Acknowledgements**
This work was funded by NSF grant NSEC DMR04-25780 to the Nano/Bio Interface Center and NIH grants (DED). AEXB is partially supported by a scholarship from the Natural Sciences and Engineering Research Council of Canada. We also thank Andrea Rehfeldt (http://www.az-design.de/) for the schematic in Fig. 6, John Beausang for help with single molecule measurements, and Klara Stefflova, Florian Rehfeldt, and Irena Ivanovska for critical reading of the manuscript.

**Appendix A**
*TIRF Excitation Scaling Correction*

The electric field intensity in TIRF decays exponentially from the sample surface. The corresponding decay of the fluorescence intensity from a dye near the surface is more complicated due its interaction with the surface, as discussed in detail in ref. (*30*). However, while these corrections can be important for comparing the intensity of a fluorophore near a surface with one far away from the surface, the corrections vary slowly with height within the first 10 nm, which is the relevant scale for our measurements. This was confirmed by measuring the drop in intensity from fluorophores on the tip of an AFM cantilever as it was moved away from the sample surface (inset **Fig. 7b**). The intensity is fit well by a single exponential with a decay length of 140 nm. This exponential decay in fluorescence intensity introduces a deviation from linear intensity scaling even for a filament that is a perfect cylinder with fluorophores arranged only around its circumference. However, for reasonable assumptions for the filament properties in our experiments, this correction is negligible for the case of fluorophores distributed in a shell. To see this, consider the geometry shown in **Fig. 7a**. It depicts a filament cross section of radius r sitting on a surface at z = 0, where z is the height above the surface. The fluorescence intensity from this cross-section is then proportional to the integrated contributions of the fluorophores around the edge times the excitation intensity at each height:

$$I_{fl} \propto \int_0^{2r} \exp[-z/d] l_{arc} dz$$

$$\propto \int_0^{2r} \exp[-z/d] (2r \cos^{-1}[\frac{r-z}{r}]) dz,$$

where $l_{arc}$ is the length of the arc between points 1 and 2 in Fig. 7, $I_{fl}$ is the fluorescence intensity, and $d$ is the characteristic decay length of the evanescent field. Using a decay length of 150 nm, the above expression for $I_{fl}$ was integrated numerically in Mathematica (Wolfram Research, Champagne, IL) for several values of filament radius up to 20 nm. As shown by the linear fit in Fig. 7b the correction due to the decay of the evanescent field is negligible compared to the experimental scatter in Fig. 3a. A similar procedure can be carried out for the case when the fluorophores are assumed to fill the cylinder uniformly. In this case, rather than the arc length appearing above, we have the *area* of a segment of the circular cross-section given by

$$A_{seg} = \frac{r^2}{2}\left(2\cos^{-1}[\frac{r-z}{r}] - \sin\left[2\cos^{-1}[\frac{r-z}{r}]\right]\right).$$

The scaling resulting from this correction is still distinguishable from a linear scaling although it is no longer perfectly quadratic (Fig. 7b).

**Appendix B**
*Single Molecule Intensity Calibration*

To determine the single fluorophore intensity we used a sample of mostly singly labeled myosin monomers and found spots that bleached in one step by monitoring the intensity over time in a square region of 5 pixels on a side (**Fig. 8a**). We then fit a 2-dimensional Gaussian to the spot prior to its bleaching and used the volume of the best fit Gaussian as the spot's total intensity (**Fig. 8b**).

| Filament Number | (Myosins / Filament) by AFM | (Myosins / Filament) by TIRF |
|---|---|---|
| i | 436 | 316 |
| ii | 430 | 292 |
| iii | 59 | 70 |
| iv | 158 | 96 |
| v | 203 | 141 |

**Table 1**: Myosins per filament as estimated from AFM height profiles and from TIRF intensity profiles after calibration of single molecule fluorescence. Filaments i-iii are shown in Fig.2.

**Figure Legend**

**Figure 1**: TIRF and AFM images of self-assembled myosin filaments on PMMA coated glass in buffer with physiological salt. Myosin molecules were labeled with an average of 1.2 dyes per heavy chain and predominantly at CYS-707 in the motor domain. (**a**) The TIRF image has limited resolution both because of diffraction effects and because of the finite pixel size of the camera. The inset shows the crystal structure of the myosin motor domain with part of the converter domain. The red star indicates CYS-707. (**b**) Tapping mode AFM height image of the same region shows the elongated and tapered structure of individual myosin filaments that are now clearly resolved. The white overlay represents the edge of the thresholded TIRF image to demonstrate registration. (**c**) Cropped and scaled TIRF image from the region indicated with the black box in (b). (**d**) Re-scan of the same region by AFM reveals still higher resolution and finer structural details of the filaments.

**Figure 2**: Fluorescence intensity (and therefore number of myosin heads) correlates with height. (**a**) TIRF image. Inset shows the optical microscope's point spread function (PSF) determined by fitting a Gaussian (standard deviation 260 nm) to a single spot in a TIRF image. (**b**) Tapping mode AFM height image of the same region. The "shadows" vertically displaced above the filaments are most likely a tip artifact, but because they are several filament widths away from the principal image, this artifact is unlikely to affect the tip tracking and height profiles along the filament lengths. (**c**) Maximum height (black) and intensity (red) profiles measured along the three filaments in the images. The blue curve is the convolution of the AFM trace with the PSF to simulate broadening of fluorescence due to diffraction and to facilitate a more appropriate comparison between height and intensity. The number in the top right of each plot is the fractional asymmetry of each filament calculated assuming the filaments have a circular cross section. It is an estimate of how much more volume the filament has on one half than on the other divided by the total volume.

**Figure 3**: Height-intensity scaling and its structural implications. (**a**) Plot of intensity versus height calculated using the broadened AFM height data. The black curve is a power law fit with a resulting exponent of 1.0 ± 0.1 suggesting that the best scaling is linear; indeed, the Akaike information criterion shows a linear fit is three times more likely to account for the data than the power law fit and $3 \times 10^6$ times more likely than a quadratic fit. Linear scaling implies that the myosin heads are arranged in a shell around the filament: if this were not the case, the number of dyes present in a given diffraction-limited cross section would scale as a higher power of the diameter (and therefore height) as illustrated schematically in (**b**).

**Figure 4**: Fractional asymmetry from raw AFM height profiles as a function of length. Black points are averages over 500 nm bins of the individual data points shown in gray. Red points correspond to the three filaments imaged in figure 2. There is large scatter about the mean of 0.32 shown by the horizontal line, but there is no clear dependence on filament length. The right panel is a histogram showing the probability distribution of the filament asymmetries, δ. Small asymmetries are the most likely (the maximum of the distribution occurs at the smallest bin), but the decay with δ is slow (approximately linear).

**Figure 5**: Pyramidal lattice model of myosin filaments. Myosin filaments are modeled using a lattice with staggered sites without allowing overhanging edges. The problem of counting the number of ways of stacking bricks on this lattice is equivalent to counting the number of "Dyck paths" (**a**). Two model filaments are shown in (**b**). To see the equivalence of these two problems, note that the path drawn behind the first filament is the same as that shown in (a), but rotated by $3\pi/4$. The base lengths *L* and fractional asymmetries δ are written in the insets. The height profiles of these filaments are equivalent to so called Dyck paths as shown by the overlay on the first filament. (**c**) Plot of δ as a function base length. Points show all possible model filaments of a given length. The red line shows the average δ. (**d**) Histogram of the points plotted in (b) showing dependence of *P*(δ) on δ.

**Figure 6**: Asymmetric myosin filaments could be motile as shown in this schematic of an asymmetric myosin filament at an F-actin junction. If the myosin filament (red) is bipolar about its center and the actin filaments (blue, minus ends towards center) are relatively immobile, then the contractile forces exerted by the myosin filament on the actin filaments will not balance. If there is sufficient asymmetry, the myosin filament will move towards the side with more heads (in this case to the right).

**Figure 7** (Appendix): (**a**) Geometry for scaling correction calculation. The region with fluorophores for the shell model is the arc of the circle between the points labeled 1 and 2. The region with flurophores for the solid model is the segment indicated by the thin lines. (**b**) The results of the numerical integration for the shell model (black points) and solid model (blue points) taking into account the decay of the TIRF field. The solid lines are fits to the data - linear in the case of the shell model and quadratic (no constant or linear term) in the case of the solid model.

**Figure 8** (Appendix): (**a**) Average intensity over time for a 5 by 5 pixel square showing a one-step bleaching event. (**b**) A Gaussian fit to a spot that subsequently bleached in a single step to determine that spot's total intensity.

Figure 1

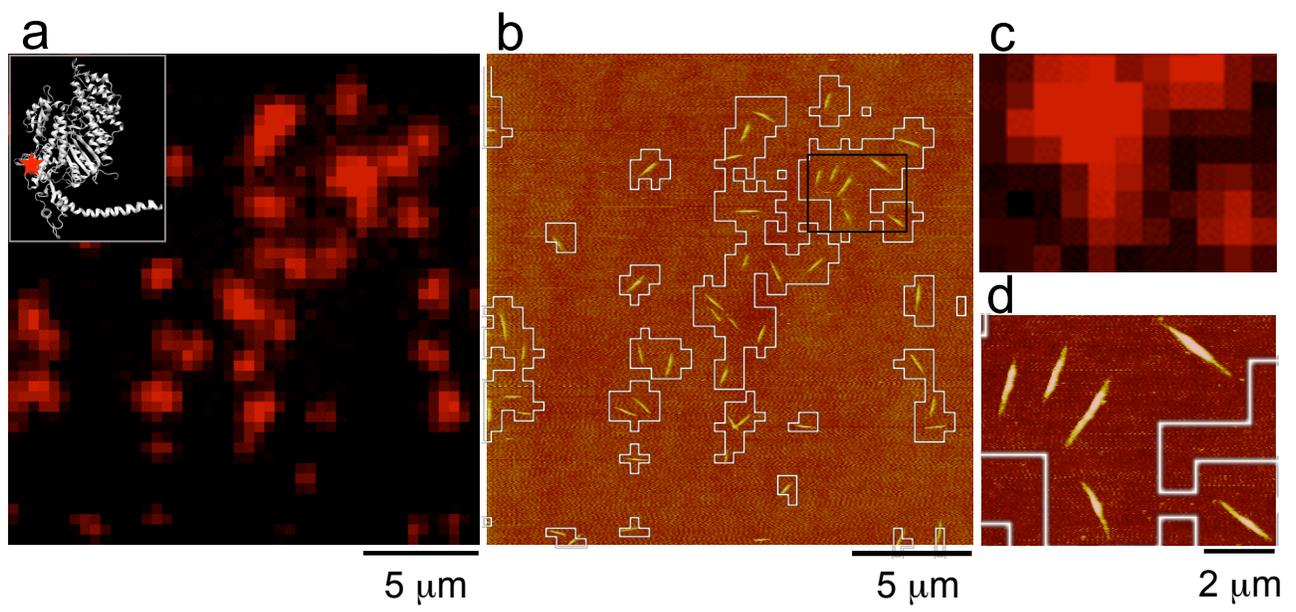

Figure 2

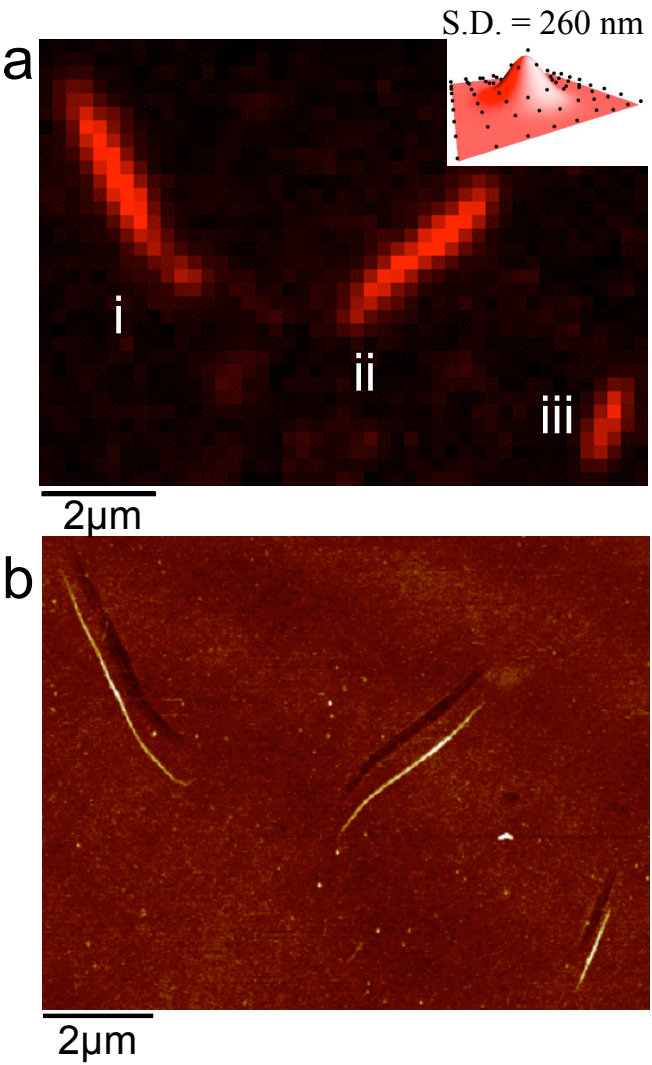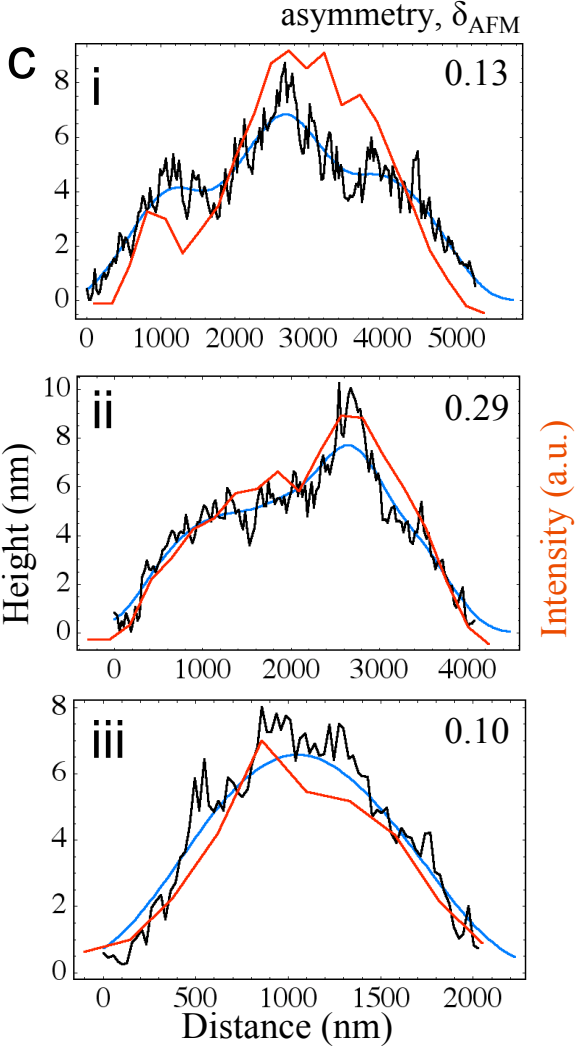

Figure 3

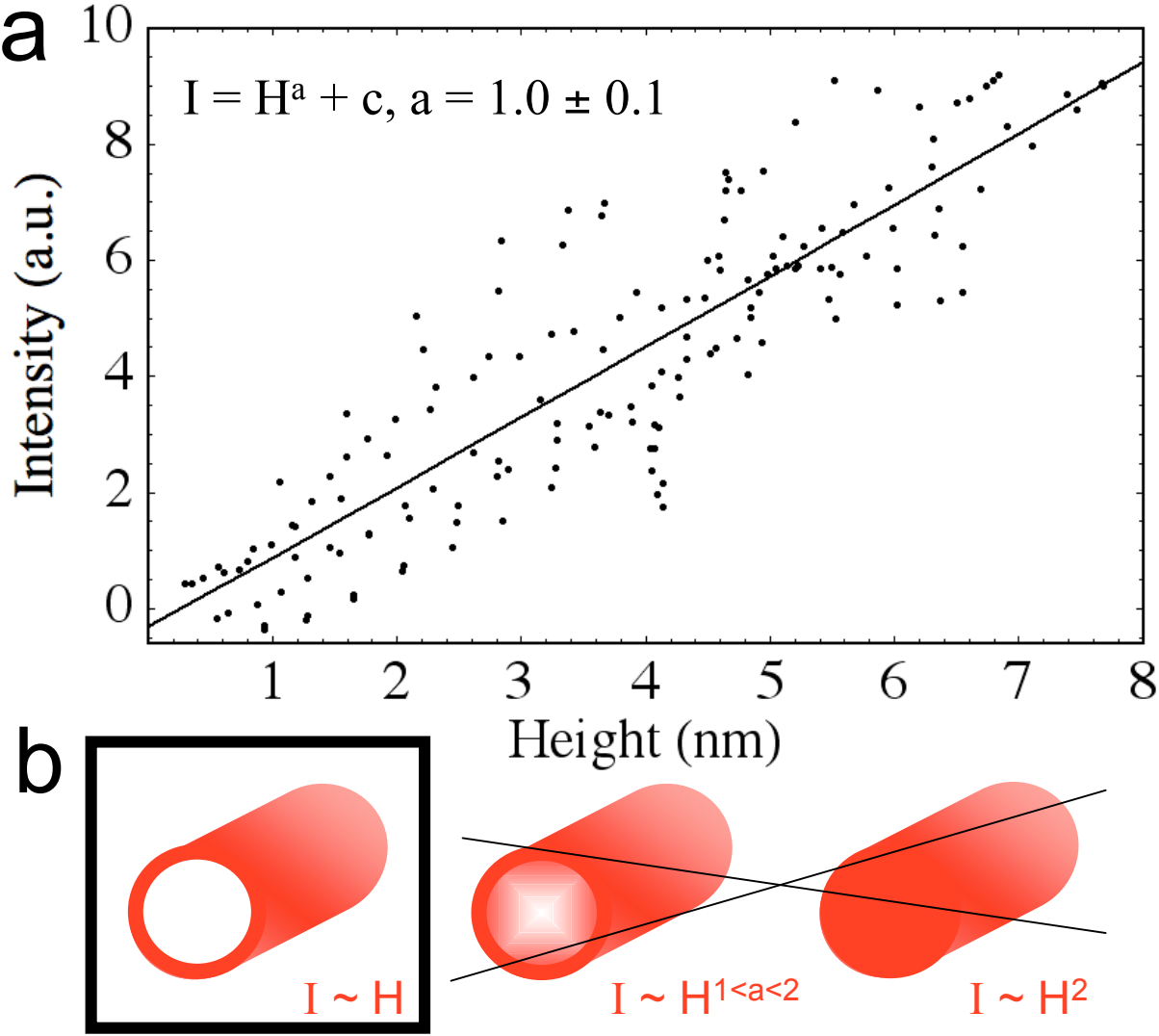

Figure 4

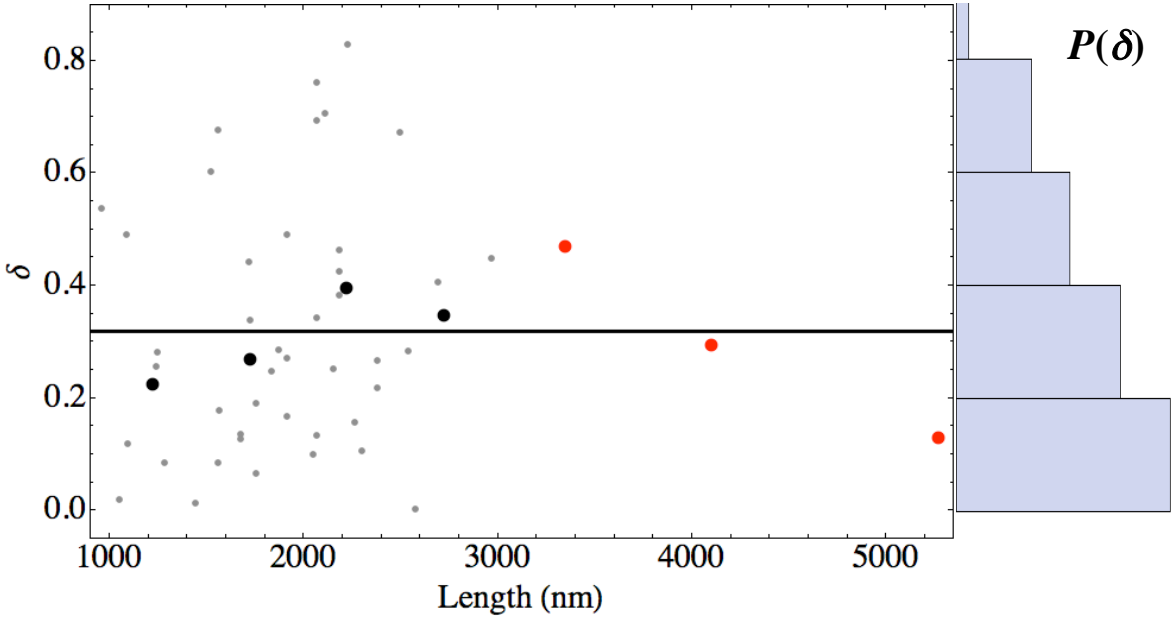

# Figure 5

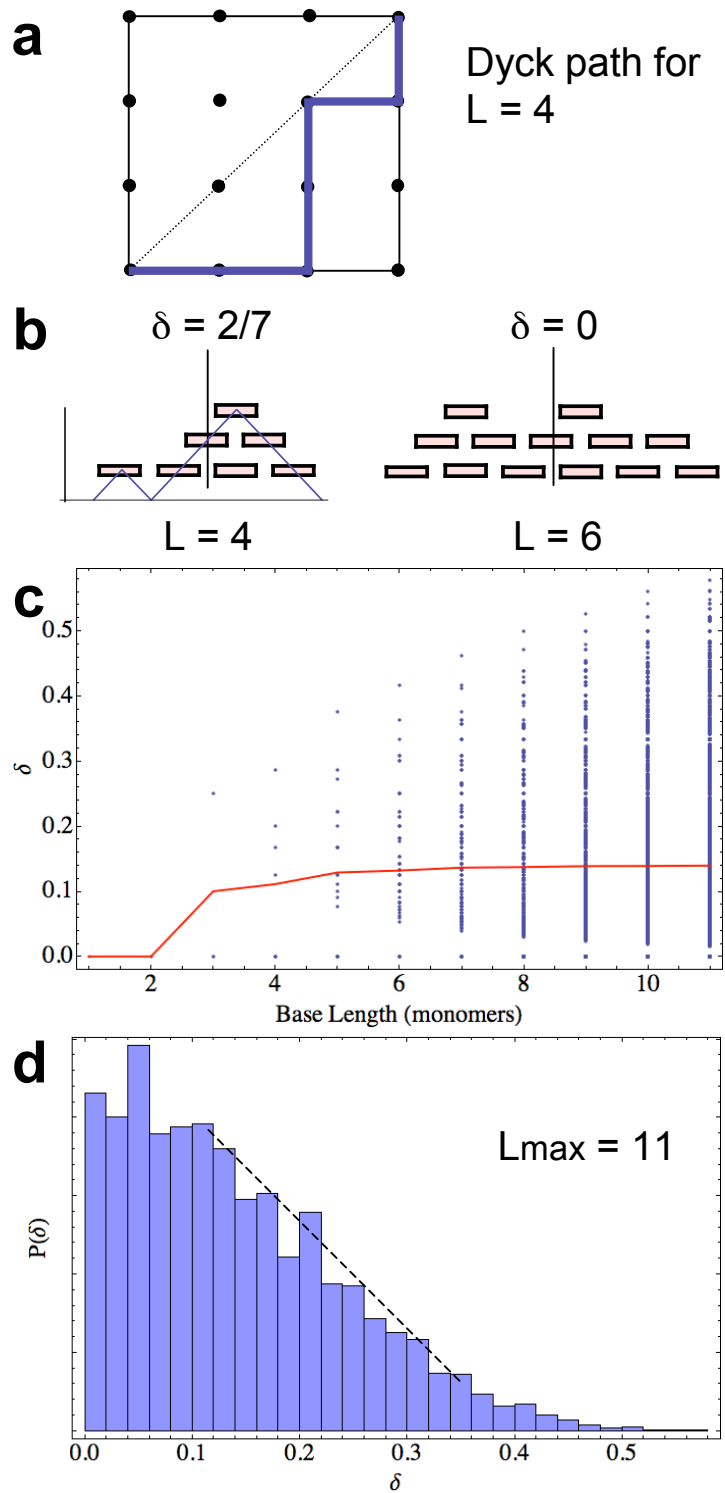

Figure 6

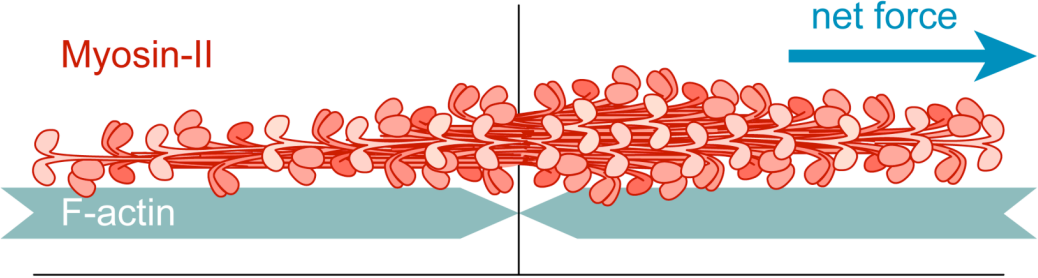

Figure 7

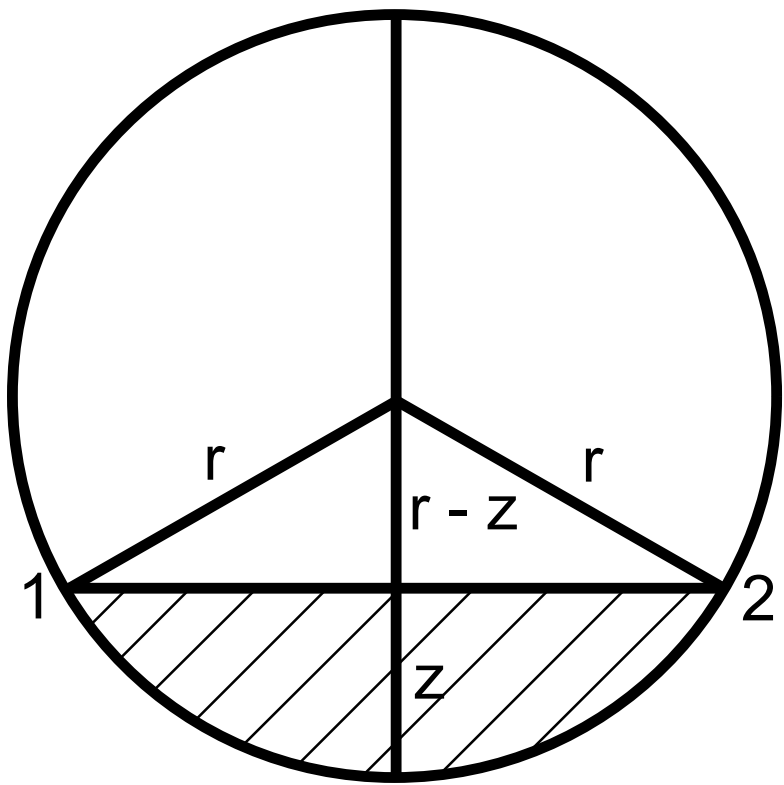

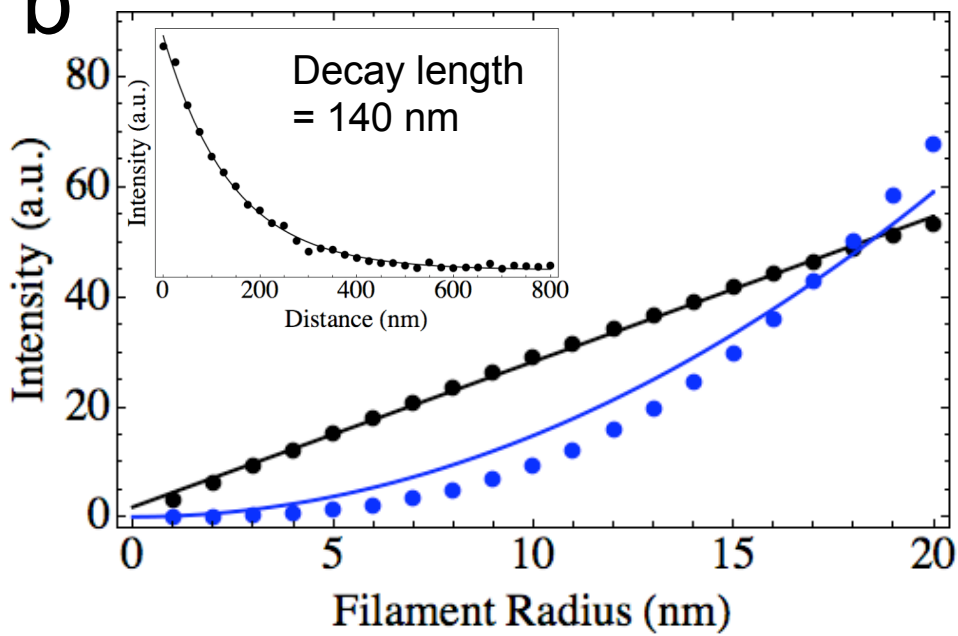

Figure 8

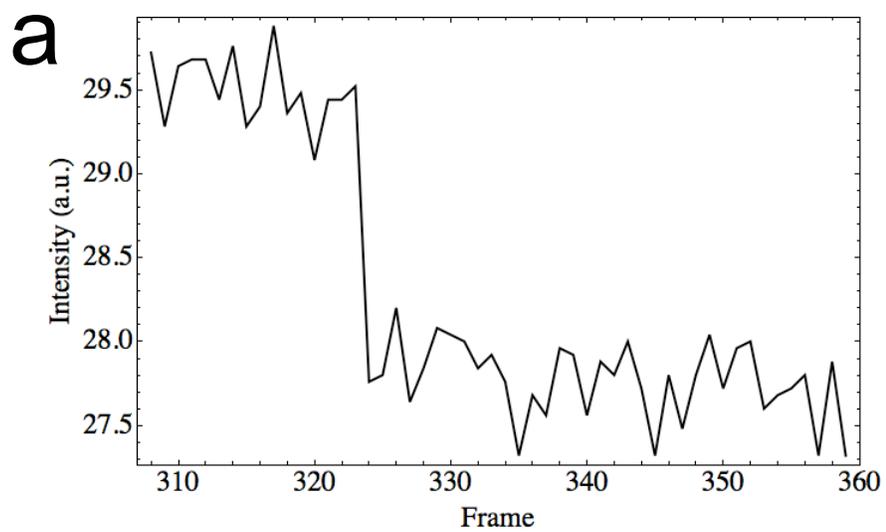

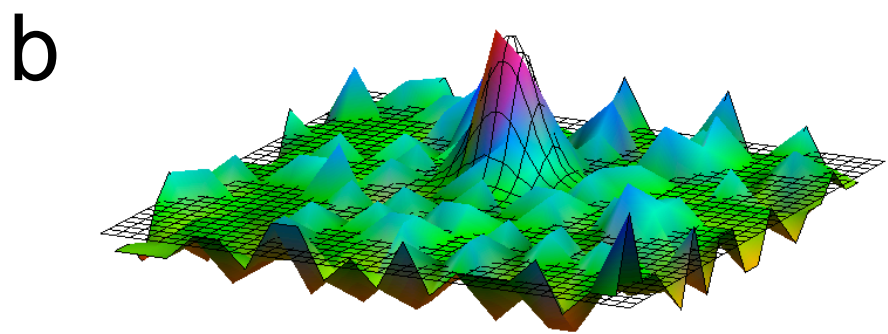